\documentclass{aastex}
\usepackage{hyperref}
\usepackage{graphicx}
\usepackage{color}
\usepackage{txfonts}
\usepackage{url}
\usepackage{natbib}
\bibpunct{(}{)}{;}{a}{}{,}
\usepackage[utf8]{inputenc}
\usepackage{subfigure}
\usepackage{verbatim}
\usepackage{amssymb}

\begin{document}

   \title{Diffusion of magnetic elements in a supergranular cell}
   \author{F. Giannattasio$^{1}$, M. Stangalini$^{2}$, F. Berrilli$^{1}$, D. Del Moro$^{1}$, L. Bellot Rubio$^{3}$}
   \affil{$^{1}$Dipartimento di Fisica, Università di Roma "Tor Vergata"\\ 
   Via della Ricerca Scientifica,1 00133 Rome, Italy\\
   $^{2}$INAF-Osservatorio Astronomico di Roma\\ 
   00040 Monte Porzio Catone (RM), Italy\\
   $^{3}$Instituto de Astrof\'isica de Andaluc\'ia (CSIC)\\ 
   Apdo. de Correos 3004, E-18080 Granada, Spain}
   \email{Fabio.Giannattasio@roma2.infn.it}

             

 
  \begin{abstract}
  Small scale magnetic fields (magnetic elements) are ubiquitous in the solar photosphere.
  Their interaction can provide energy to the upper atmospheric layers, and contribute to heat the solar corona.
  In this work, the dynamic properties of magnetic elements in the quiet Sun are investigated.
  The high number of magnetic elements detected in a supegranular cell allowed us to compute their displacement spectrum $\langle(\Delta r)^2\rangle\propto\tau^\gamma$ (being $\gamma>0$, and $\tau$ the time since the first detection), separating the contribution of the network (NW) and the internetwork (IN) regions.
  In particular, we found $\gamma=1.27\pm0.05$ and $\gamma=1.08\pm0.11$ in NW (at smaller and larger scales, respectively), and $\gamma=1.44\pm0.08$ in IN.
  These results are discussed in light of the literature on the topic, as well as the implications for the build up of the magnetic network.
  
  \end{abstract}

   \shorttitle{Diffusion of magnetic elements in supergranular cells}
   \shortauthors{F. Giannattasio et al.}

   \maketitle
%
\section{Introduction}
Magnetic field is ubiquitous in the quiet solar photosphere and interacts with plasma at all scales, from granular and mesogranular scales \citep[see, e.g.,][]{1980PhDT.........3N, 1998A&A...330.1136R, 2004A&A...428.1007D, 2005ApJ...632..677B,2007ApJ...666L.137C, 2011ApJ...727L..30Y, 2013SoPh..282..379B} to supergranular and giant scales \citep[see, e.g.,][]{1956MNRAS.116...38H, 1964ApJ...140.1120S, 2004ApJ...616.1242D, 2007A&A...472..599D, 2008ApJ...684.1469D, 2012ApJ...757...19B, 2012ApJ...751....2O, 2012ApJ...758L..38O, 2014ApJ...784L..32M}.
Such an interaction may give rise to magnetic reconnections and the excitation of magnetohydrodynamic (MHD) waves, which are the main mechanisms proposed to explain the heating of the solar corona \citep[see, e.g.,][]{1947MNRAS.107..211A, 1988ApJ...330..474P, 2003PhRvL..90m1101H, 2007Sci...318.1574D, 2007Sci...317.1192T, 2011ApJ...736....3V, 2013A&A...554A.115S, 2013A&A...559A..88S}.
Therefore, it is important to investigate the interaction between small-scale magnetic fields (hereafter magnetic elements) in the quiet Sun \citep[see, e.g.,][]{2003PhRvL..90m1101H, 2006ApJ...652.1734V}, and between magnetic elements and plasma flows.
\citet{2012ApJ...758L..38O} showed that magnetic elements are dispersed radially in supergranules with velocities aligned with the plasma flow.

It has been shown that the efficiency with which the magnetic elements are transported on the photosphere is well described by a power law \citep[see, e.g.,][]{2013ApJ...770L..36G} $\langle(\Delta r)^2\rangle=c\tau^\gamma$, where $\langle(\Delta r)^2\rangle$ is the mean square displacement, $c$ a constant, $\tau$ is the time measured since the first detection, and $\gamma$ is the spectral index.
This parameter quantifies the efficiency of the transport process with respect to a random walk (RW, or normal diffusion, for which $\gamma=1$).
When $\gamma\neq1$ the process is termed anomalous diffusion.
In particular, the regime $\gamma>1$ corresponds to a super-diffusion, while $\gamma<1$ to a sub-diffusion.
Recently, several authors, both using G-band images and magnetograms from ground and space observations, agreed on a super-diffusive dynamic regime \citep[see, e.g.,][]{2001PhRvL..86.5894L, 2011A&A...531L...9M, 2011ApJ...743..133A, 2013ApJ...770L..36G, 2014arXiv1401.7522J}.
In these works, magnetic elements are assumed to be passively transported by the flow. 

The aim of this paper is to determine the spectral index and the diffusivity corresponding to magnetic elements within a supergranular cell, separating the contributions from network (hereafter NW) and internetwork (hereafter IN) regions, where different MHD regimes are expected.
For this purpose, we take advantage of a very long ($25$ hours without interruption) time sequence of high resolution magnetograms, acquired in the quiet Sun by the Solar Optical Telescope (SOT) onboard Hinode.

\section{Observations and analysis}
\label{Section:Obs_and_analysis}
\label{Section:Results}
\begin{figure}[ht!]
 \subfigure[][Horizontal velocity field]{\includegraphics[width=8cm]{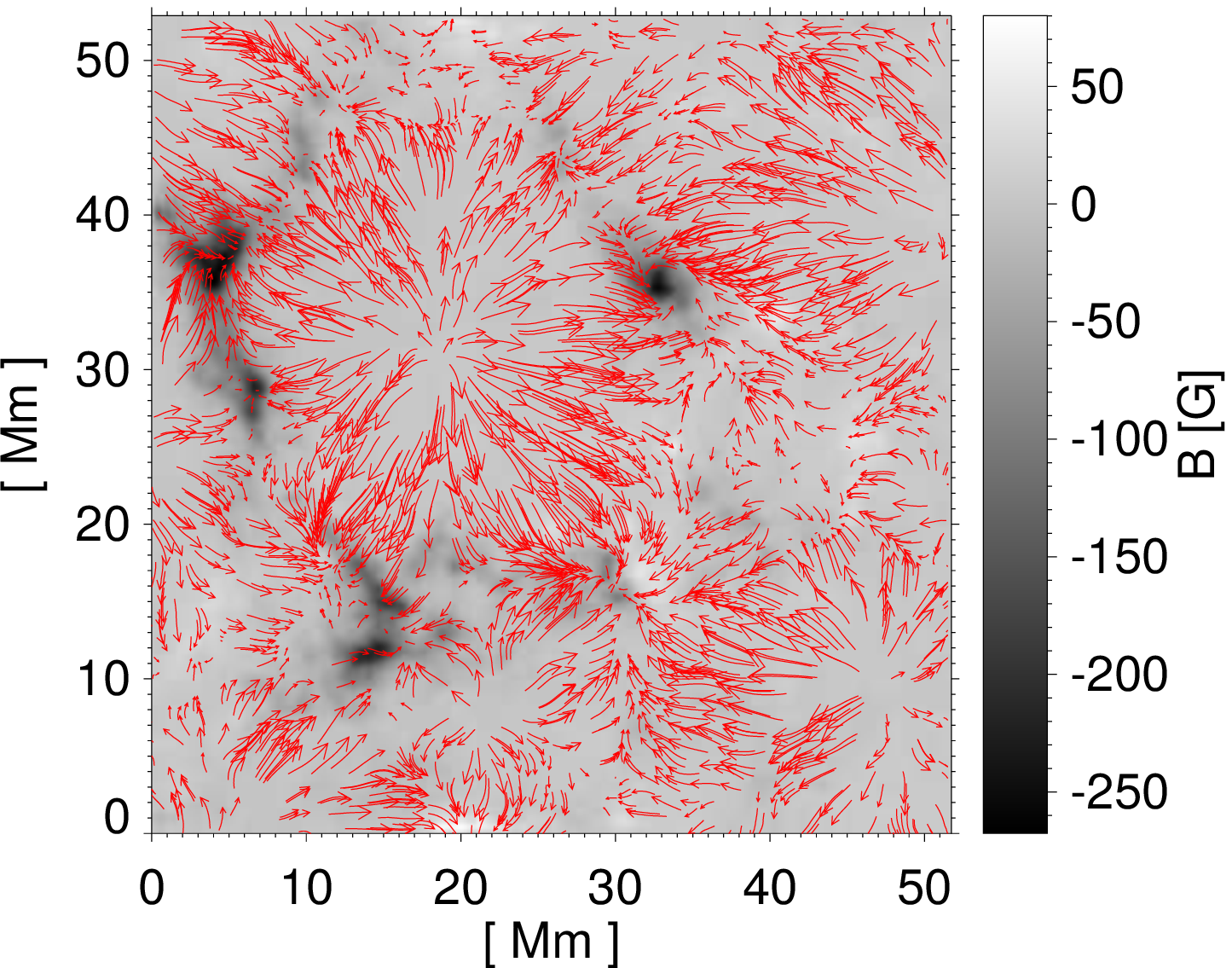}}
 \subfigure[][Horizontal velocity amplitude]{\includegraphics[width=8cm]{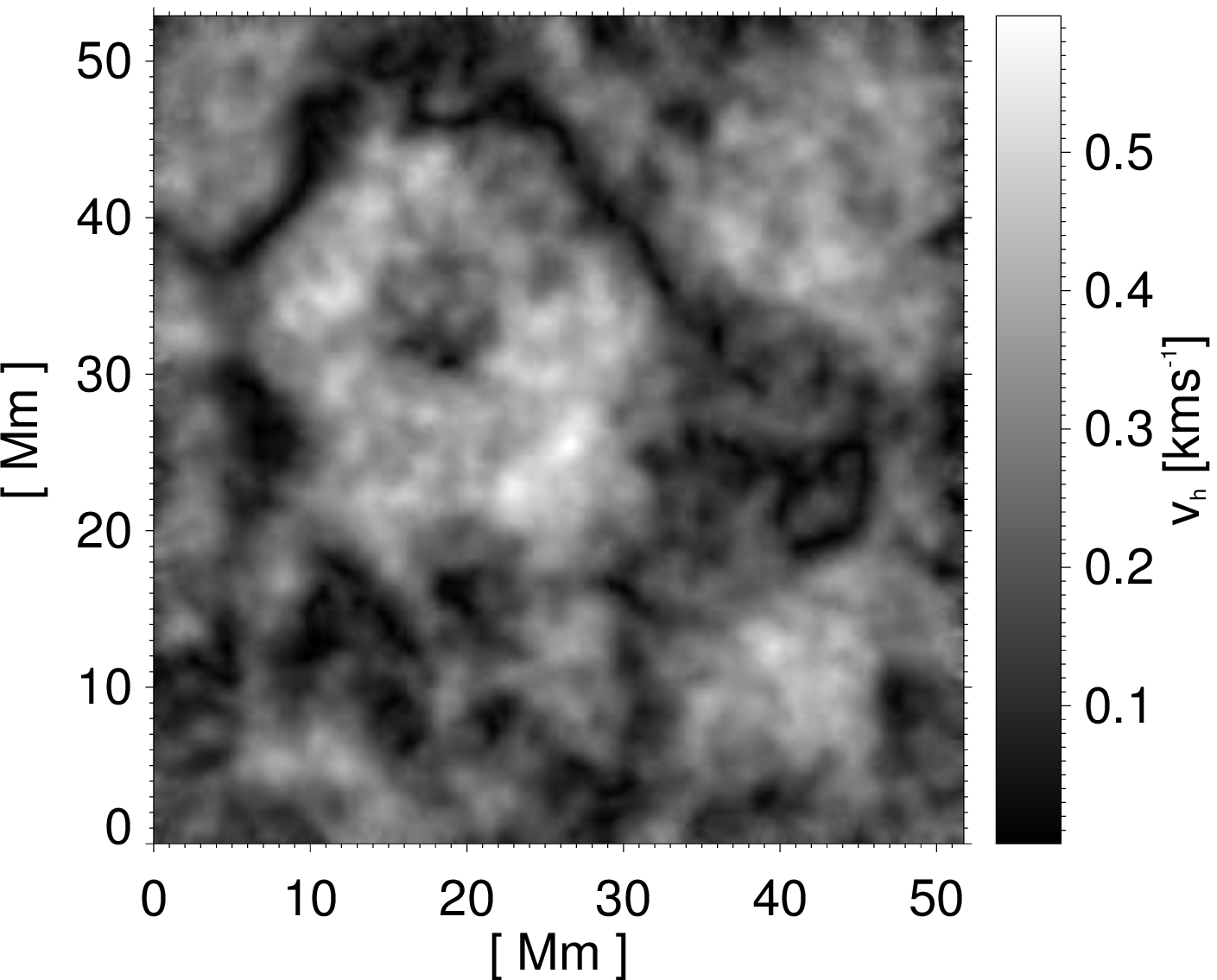}}
 \subfigure[][Saturated mean magnetogram]{\includegraphics[width=8cm]{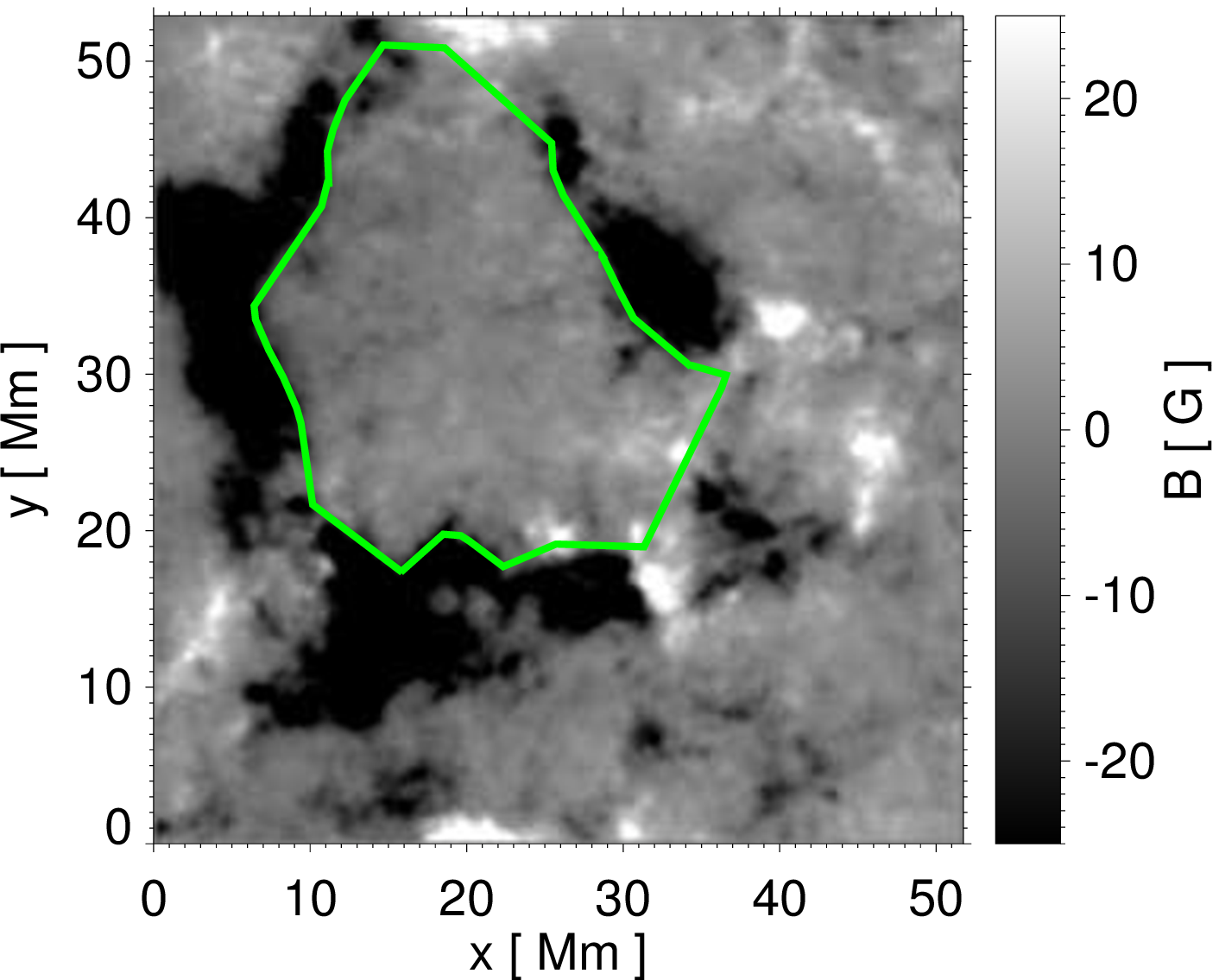}} 
 \subfigure[][          Deep magnetogram]{\includegraphics[width=8cm]{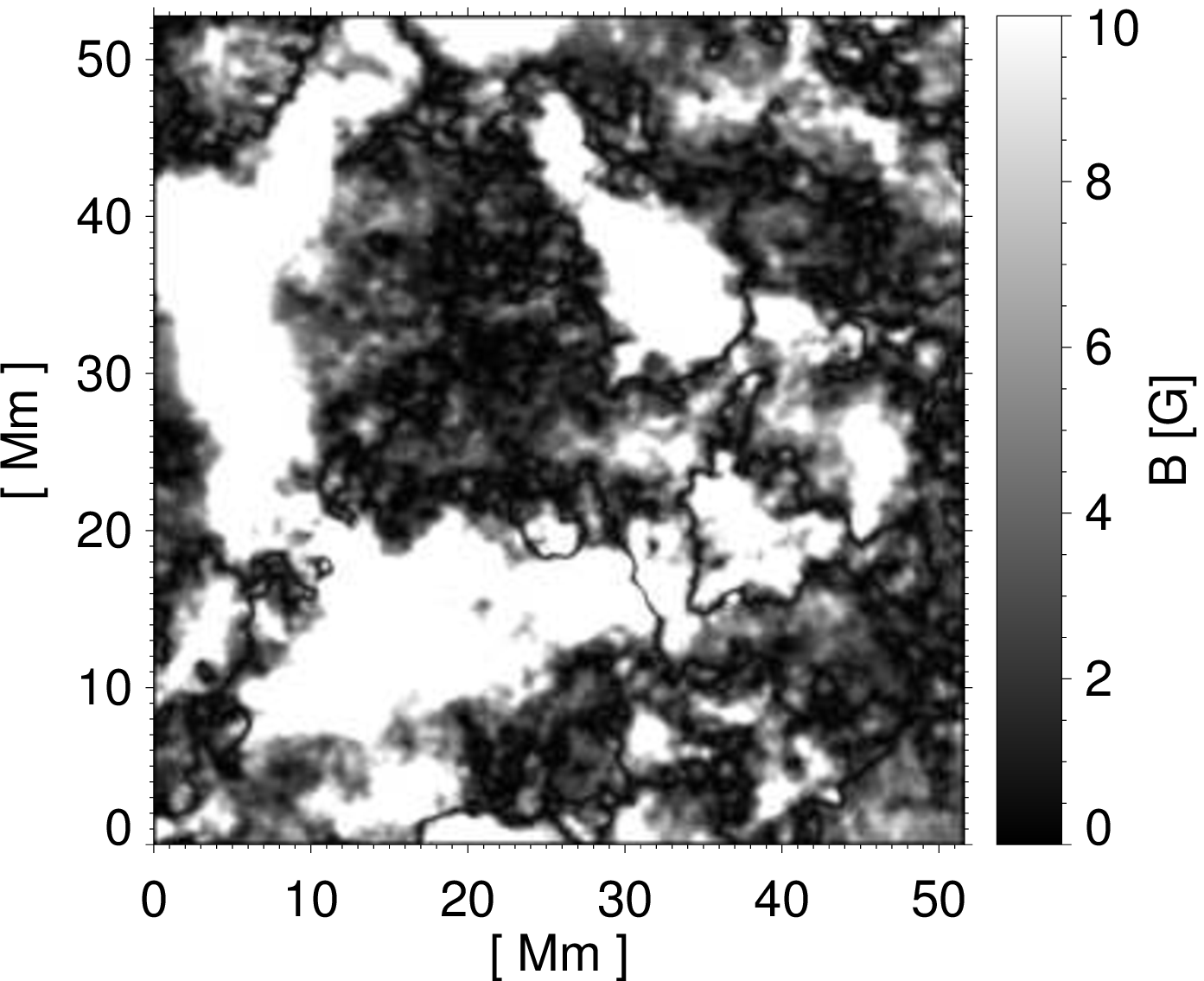}}
 
   \caption{Top row: 25-hr averaged horizontal velocity field, as computed by the LCT technique, in direction (red arrows in panel a) and amplitude (panel b). The 25-hr averaged magnetogram is shown in the background on panel (a). Bottom row: average magnetogram saturated at $25$ (panel c) and $10$ G (panel d). The green polygon in panel (c) highlights the supegranular cell.}
   \label{Fig:Mags}
\end{figure}
In this work we analyzed the same Hinode/SOT data \citep{2007SoPh..243....3K, 2008SoPh..249..167T} described in \citet{Milan} and used by \citet{2013ApJ...770L..36G} to perform their analysis on the diffusion of magnetic elements up to supergranular scales.
These data belong to the {\itshape Hinode Operation Plan 151} (HOP151) entitled {\itshape Flux replacement in the photospheric network and internetwork}, and consist of a series of magnetograms and filtergrams $25$ hr long, acquired by the {\itshape Hinode} Narrowband Filter Imager (NFI) in a quiet-Sun region near disk center.
The series starts at 08:00:42 on 2010 November 2.
The magnetogram noise level is $\sigma=6$ G for individual frames.
Observations were performed at two wavelengths $\pm160$ m\AA\ from the core of the NaI D spectral line at $589.6$ nm.
The pixel size is $0".16$ ($116$ km on the solar photosphere), corresponding to a spatial resolution of $0".32$.
The field of view (FoV) area is $\sim51\times53$ Mm$^2$.
The cadence between successive frames (a total of $995$) is $90$ s.
Five-minute oscillations were filtered out by using a $k-\omega$ filter with $c_s=7$ kms$^{-1}$.
All the details regarding the calibration of data can be found in \citet{Milan}.

In order to compute the horizontal velocity field, we used the {\itshape Fast Local Correlation Tracking} code  \citep[FLCT;][]{2008ASPC..383..373F}, with a spatial window of 10 pixels ($\sim1200$ km).
All the $995$ velocity frames obtained were summed up to recover the 25-hr averaged horizontal velocity field. 
The results of the LCT analysis are shown in Figure \ref{Fig:Mags}.
In that figure, the red arrows in panel (a) mark the retrieved average horizontal velocity field.
In the background of the same panel we show the 25-hr averaged magnetogram.
The amplitude of the horizontal velocity field is displayed in panel (b).
Both the mean magnetogram and the horizontal velocity field unambiguously show a whole supergranular cell in the FoV (highlighted by the green polygon in panel (c)).

The iterative procedure described in \citet{2004A&A...428.1007D} and used in \citet{2013ApJ...770L..36G} was then applied on the magnetograms to detect and track the magnetic elements.
To avoid false detections due to time-uncorrelated signals, only those lasting at least $5$ frames were considered.
A total of $20145$ magnetic elements was tracked in $25$ hr. 
Their lifetime spans the range between $7.5$ minutes (set by the selection criterion) and $11.1$ hr.

As aforementioned, we were interested in the estimation of the diffusivity of the advected magnetic elements in a supergranular cell. 
For this purpose, among all the magnetic elements we considered two subsamples: i) those belonging to NW; ii) those belonging to IN.
NW and IN were defined from the average magnetogram.
In panel (c) of Figure \ref{Fig:Mags} we show the average magnetogram saturated at $B_s=25$ G ($4$ times the noise level).
The pixels adjacent to the polygon of panel (c) with $|\mathbf{B}|\geq B_s$ were defined to belong to NW.
The pixels within the polygon and with $|\mathbf{B}|<B_s$ were defined to belong to IN.
We found $2021$ magnetic elements which are first detected, live and die within the NW; and $3563$ magnetic elements which are first detected, live and die within the IN, respectively. 
For these subsamples we computed the mean square displacement as a function of time, $\langle(\Delta r)^2\rangle(\tau)$.
The log-log plot of such a function, namely the displacement spectrum, consists in a line with slope $\gamma$.
The diffusivities were retrieved by the relations \citep{1975mit..bookR....M, 2011ApJ...743..133A}
$$
K(\Delta r)=\frac{c\gamma}{4}\left(\frac{\Delta r^2}{c}\right)^{\frac{\gamma-1}{\gamma}},
$$
$$
K(\tau)=\frac{c\gamma}{4}\tau^{\gamma-1},
$$
which describe the anomalous diffusion processes occurring in the photosphere as a function of both the spatial ($\Delta r$) and temporal ($\tau$) scales.
Together with $\gamma$, also the constant $c$ was retrieved by fitting the displacement spectrum.

\section{Results}


Our data set allowed us to study the dynamic properties of the magnetic elements on a large range of spatial and temporal scales; from granular to supergranular ones.
Moreover, since the FoV includes a whole supergranule, it was possible to study separately the dynamics of the NW magnetic elements and that of the IN ones in a supergranular environment.

Figure \ref{Fig:Mags} shows that, on average, the horizontal velocity field inside the supergranular cell is radial, and directed from the center to the boundaries (panel (a)).
The horizontal velocity is of order $\sim10$ ms$^{-1}$ in the center of the cell.
Moving outwards, it reaches a maximum of $\sim600$ ms$^{-1}$ at about half the cell radius, then decreases again to very low values at the boundaries (panel (b)).
These results are consistent with those found by \citet{2012ApJ...758L..38O}, who analyzed another data set belonging to HOP151.

A low magnetic flux ($\lesssim5$ G) is observed in the IN, where magnetic elements are weak and highly dynamic, standing in the same region for a shorter time than those within the NW.
In order to increase the image contrast and inspect the regions with enhanced mean magnetic flux, we saturated the mean magnetogram at $10$ G, obtaining the deep magnetogram in the FoV shown in panel (d) of Figure \ref{Fig:Mags}.
The deep magnetogram also confirms the findings by \citet{2012ApJ...755..175M, 2013SSRv..178..141M, 2014A&A...561L...6S}, by showing a deficit of magnetic flux at the center of the supergranule at those locations where the horizontal velocity is slower.
When focusing on the IN, radial structures appear which, under the hypothesis that magnetic elements are passively transported by the flow, retrace the average horizontal velocity field lines computed by LCT.
In the NW, magnetic elements are stronger, more compact and larger in size.
Due to the joint action of their augmented magnetic flux (which allows them to withstand the plasma drag force) and the slower horizontal velocity fields therein, they remain in the same regions for longer. 
The NW evolution takes place on several generations of elements, comparable with the supergranular time scales (about one day). 
On the contrary, IN magnetic elements are weaker and tend to not stay for long in the same regions.

Such a different behavior should influence the efficiency with which the magnetic field is carried along in a supergranular cell. 
This is important because it affects the rate of magnetic field interactions, which is a fundamental parameter to evaluate the energy with which the quiet Sun contributes to heat the upper atmosphere.
To investigate this point, we computed the displacement spectrum of NW and IN magnetic elements separately.
\begin{figure*}[ht!]
 \resizebox{\hsize}{!}{\includegraphics{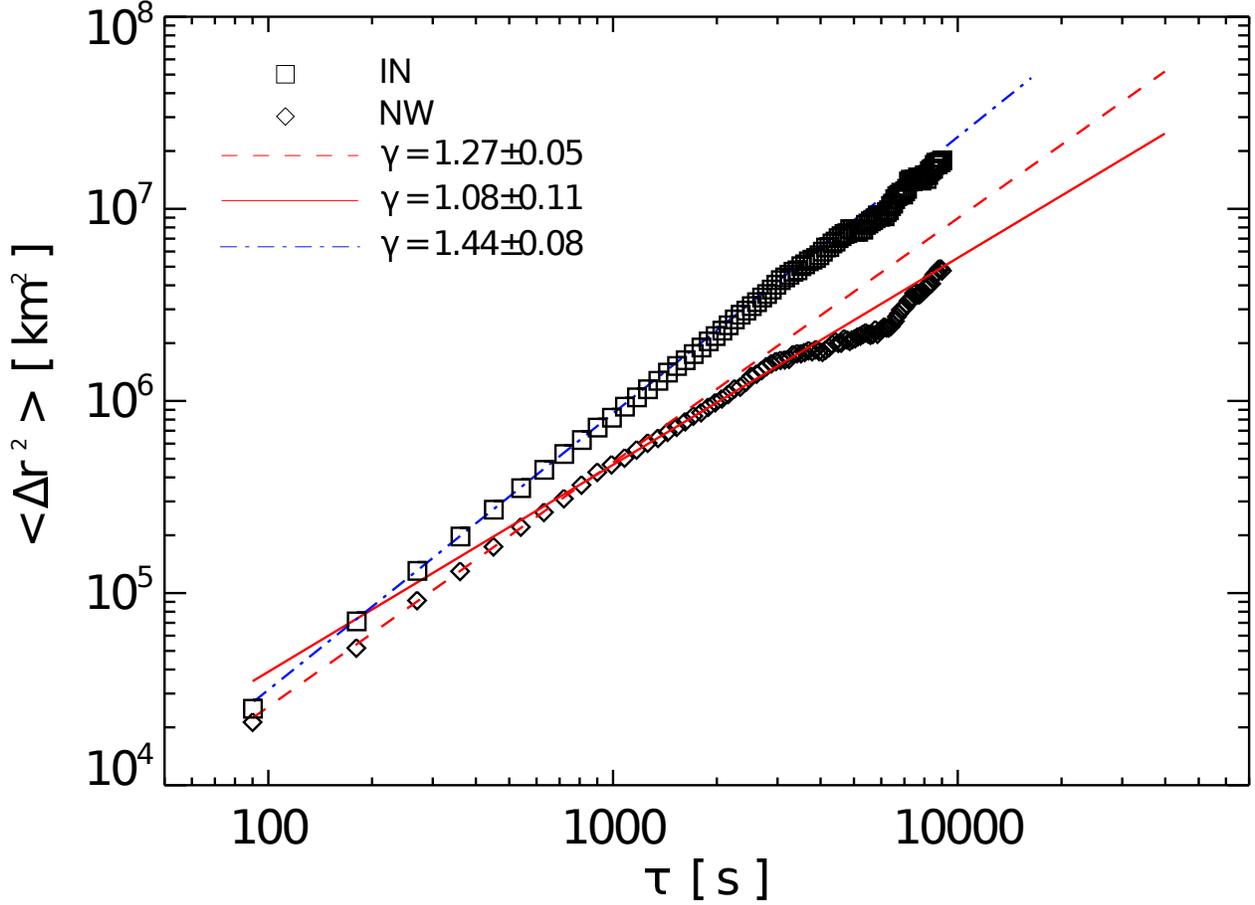}}
 \caption{Displacement spectrum for IN (black squares) and NW (black diamonds) magnetic elements. The blue dash-dotted line fits the IN data points. The red lines fit the NW data points for $\tau\lesssim600$ s (dashed line) and $\tau\gtrsim600$ (solid line).}
   \label{Fig:gamma}
\end{figure*}

\begin{figure*}[ht!]
\centering
   \subfigure[][]{\includegraphics[width=7.5cm]{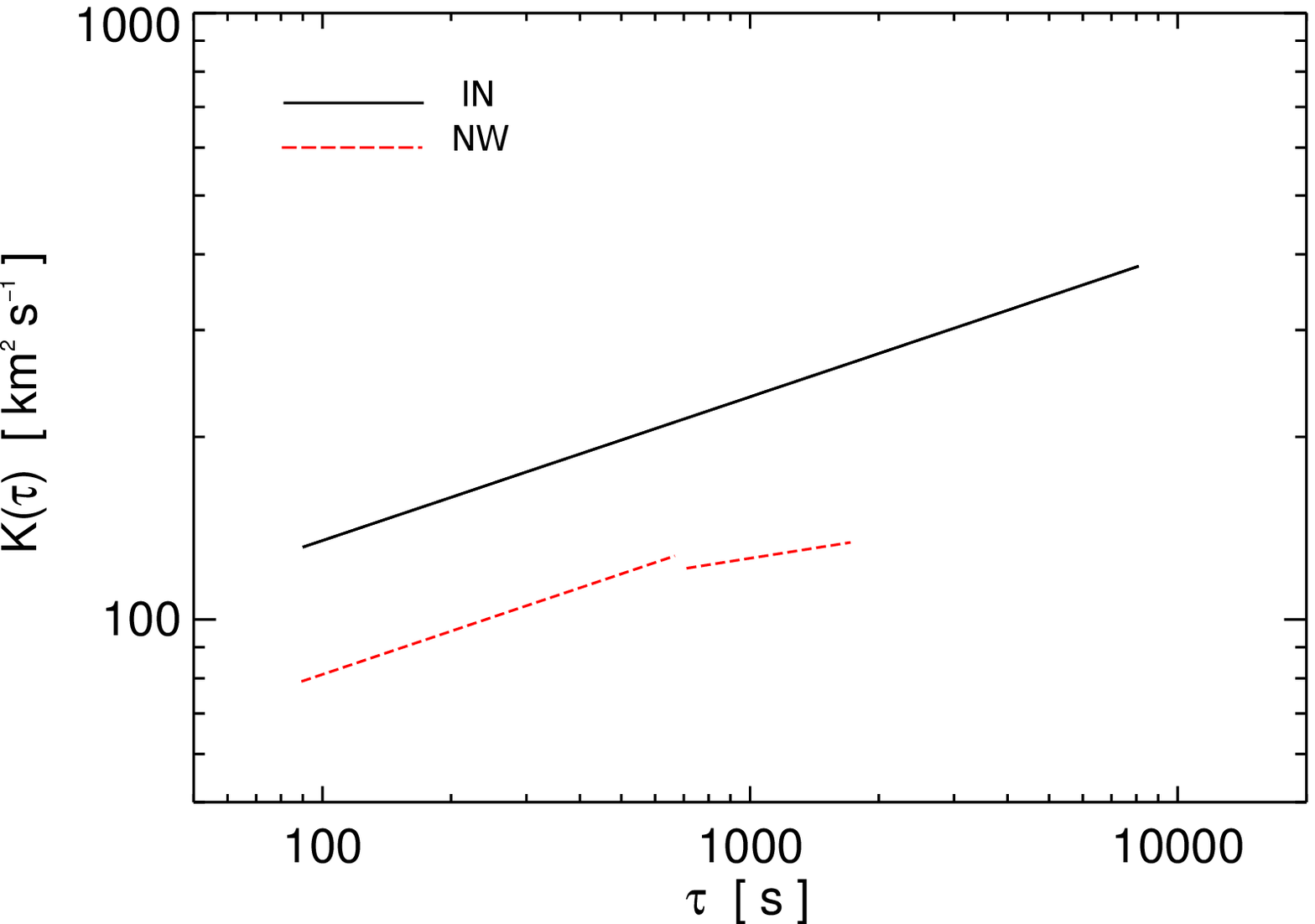}}
   \subfigure[][]{\includegraphics[width=8cm]{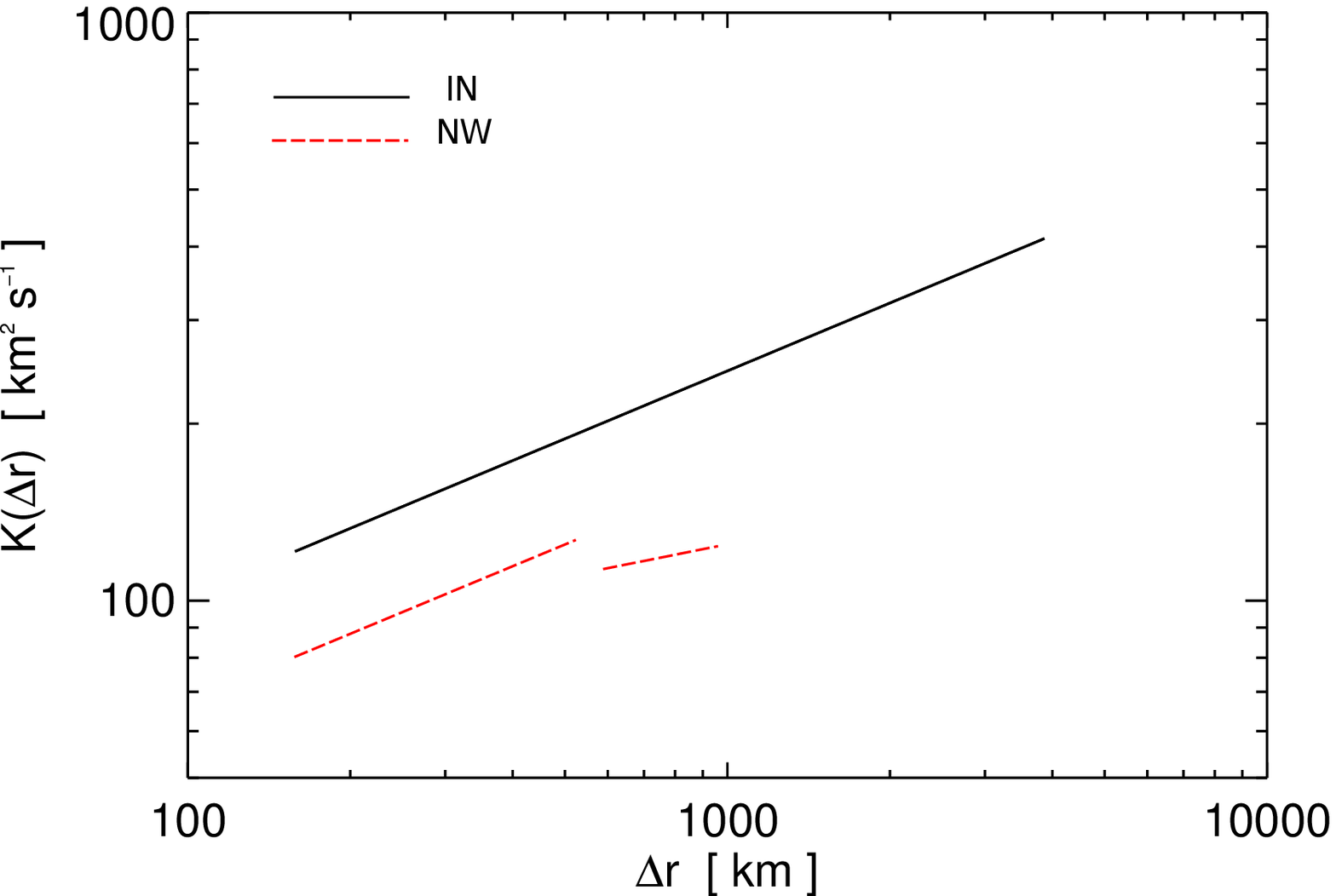}}
   \caption{Diffusivity as a function of the temporal (panel a) and the spatial scale (panel b) for IN (black solid line) and NW (red dashed line) magnetic elements.}
   \label{Fig:Diffusivity}
\end{figure*}
The results are shown in Figure \ref{Fig:gamma}.
In the IN (square symbols), the spectrum can be fitted with a single-slope power law (the blue dash-dotted line). The retrieved spectral index is $\gamma=1.44\pm0.08$, which corresponds to a super-diffusive regime.
The uncertainties on $\gamma$ were computed as the standard deviations of the distributions obtained by randomly subsampling the magnetic elements.
In the NW (diamond symbols), two power laws were used, according to the statistical method described in \citet{1999GeoRL..26.2801M}.
We found a spectral index $\gamma=1.27\pm0.05$ (a super-diffusion) for temporal (spatial) scales $\tau\lesssim600$ s ($\Delta r\lesssim500$ km, the red dashed line), and $\gamma=1.08\pm0.11$ (which is consistent with a RW) for $\tau\gtrsim600$ s ($\Delta r\gtrsim500$ km, the red solid line).
A double slope regime was first observed by \citet{2013ApJ...770L..36G} performing the same analysis on the same data set, but considering the whole FoV.
The comparison with that work indicates that at small scales the IN magnetic elements give the greatest contribution to the magnetic field dynamics in the quiet Sun, while NW elements dominate at higher scales and are responsible for the change of slope.

A possible scenario to interpret these results for a superganular cell is the following. 
In the IN, weak magnetic elements are transported efficiently at all scales. 
In the NW, magnetic flux enhancement lowers the transport rate on small scales, and magnetic fields fall into velocity sinks (or traps) at a certain decorrelation mean time.
From that time on, magnetic elements evolve according to the underlying velocity pattern, along more random paths. 
As we do not observe a change of slope in the IN regions, the velocity sinks should be more effective in the NW.\\
Another possible interpretation implies the distinction between the strong and weak magnetic regimes observed in a supergranular cell \citep[see][]{2003ApJ...588.1183C}.
While in IN, where a weak-field regime holds, convection effectively transports the magnetic elements, in NW, where a strong-field regime holds, convection is reduced, to a point that magnetic elements cannot be further trasported and aggregated.

In Figure \ref{Fig:Diffusivity} we show the diffusivity $K$ as a function of the temporal (panel a) and spatial scales (panel b). 
In the IN the diffusivity increases from $\sim100$ to $\sim400$ km$^2$s$^{-1}$.
In the NW the diffusivity increases from $\sim80$ to $\sim150$ km$^2$s$^{-1}$ for $\tau\lesssim600$ s ($\Delta r\lesssim500$ km); and much slower for longer temporal (spatial) scales.
The value of the diffusivity is related to the possibility to aggregate and amplify the magnetic field at a certain scale \citep{2011ApJ...743..133A}.
In the IN, the diffusivity is lower at small scales, and increases linearly with space and time due to super-diffusion. 
Then it follows that the amplification of the magnetic field is much more likely to happen on small scales, as it can resist to the spreading action of the flow.
In the NW magnetic fields amplify more easily, as the diffusivity is a bit lower.
On larger scales the difference in diffusivity between IN and NW increases more and more, especially on scales $\tau\gtrsim600$ s and $\Delta r\gtrsim500$ km.
Thus, in a supegranular cell the amplification at large scales should be facilitated in the NW.


The diffusion regimes found in NW and IN can help to address the following question.
Does the magnetic field in the quiet Sun emerge in the central regions of supergranules and evolve toward the boundaries?\\
An analytical expression for the typical radial velocity profile expected in supergranules was found by \citet{1989ApJ...345.1060S}, and successfully applied by \citet{2012ApJ...758L..38O} to HOP151 data.
It reads $v(r)=V(r/R)exp(-r^2/R^2)$, $r$ being the distance from the center of the supegranule, $R$ its radius, and $V$ a free velocity parameter.
By considering an upper limit for the typical velocity of $V=1$ kms$^{-1}$ \citep[see ][]{2012ApJ...758L..38O}, and $R\sim10$ Mm (found by visulal inspection of Figure \ref{Fig:Mags}), we can retrieve an average horizontal velocity inside the supergranule of amplitude $\bar{v}=\frac{1}{R}\int_0^Rv(r)dr\simeq0.32$ kms$^{-1}$.
This implies a crossing time (i.e., the time needed to go from the center of the supergranule to the boundaries) of $t_{cross}\gtrsim8.7$ hr.
We remark that this is a lower limit value, as we have implicitly assumed a ``ballistic'' motion with $\gamma=2$, so that $v\sim\langle\Delta r\rangle/\tau$.
According to the lifetime distribution of \citet{2013ApJ...770L..36G}, only a negligible fraction of magnetic elements ($\sim0.01\%$), would have a lifetime long enough to survive and reach the boundaries of the supergranular cell by starting from the center.
In contrast to previous literature \citep[see, e.g., ][]{2003ApJ...588.1183C}, this argument rules out the possibility for the magnetic elements which arise in the center of supergranular cells to contribute to the building up of the magnetic network. 
Our results are consistent with those shown in the recent work of \citet{2014A&A...561L...6S}, who found that the small scale flux emergence is not homogeneous over the supergranules, and tends to be less efficient at the centers of supergranules, and more efficient near the boundaries.
\section{Conclusions}
The rate of transport of magnetic elements in the quiet Sun is a fundamental parameter to constraint the energy supplied by the interactions between magnetic fields.
Such an energy contributes to heat the solar corona.
We studied the diffusion properties of magnetic elements in a supegranular cell.
By LCT analysis, we found that the amplitude of horizontal velocity field is small in the center of the supergranular cell ($\sim10$ ms$^{-1}$), and increases toward the boundaries.
A maximum is observed at about half the radius ($\sim600$ ms$^{-1}$), consistently with the most recent literature.
Under the hypothesis of magnetic elements passively transported by the flow, we found a super-diffusive dynamic regime with $\gamma=1.44\pm0.08$ in IN; while $\gamma=1.27\pm0.05$ and $\gamma=1.08\pm0.11$ in NW, at smaller and larger scales, respectively.
This confirms that different dynamic regimes are expected in supergranules.
In particular, the lower diffusivity of magnetic elements in NW regions allows to amplify more easily the magnetic fields therein.
We remark that for the first time we showed a variation of $\gamma$ with the horizontal velocity field in a supergranule.
Finally, by a simple calculation we found that the time needed for magnetic elements to cross the supergranule from its center to the magnetic network is $\sim8.7$ hr. 
Only a couple of magnetic elements have a lifetime comparable with the crossing time.
This suggests that magnetic elements emerging in the center of a supergranular cell cannot survive long enough to migrate to the boundaries and form the network.

\acknowledgements
      This work is supported by a grant at University of Rome Tor Vergata, and by the PRIN-INAF 2010 grant, funded by the Italian National Institute for Astrophysics (INAF).\\    
      This work has benefited from discussions in the Flux Emergence meetings held at ISSI, Bern in December 2011 and June 2012.\\
      The data used here were acquired in the framework of {\itshape Hinode} Operation Plan 151, entitled {\itshape Flux replacement in the solar network and internetwork}.\\
      {\itshape Hinode} is a Japanese mission developed and launched by ISAS/JAXA, collaborating with NAOJ as a domestic partner, NASA and STFC (UK) as international partners. Scientific operation of the Hinode mission is conducted by the Hinode science team organized at ISAS/JAXA. This team mainly consists of scientists from institutes in the partner countries. Support for the post-launch operation is provided by JAXA and NAOJ (Japan), STFC (U.K.), NASA, ESA, and NSC (Norway).




\end{document}